# Combined Airborne Wind and Photovoltaic Energy System for Martian Habitats

Lora Ouroumova, Daan Witte, Bart Klootwijk, Esmée Terwindt, Francesca van Marion, Dmitrij Mordasov, Fernando Corte Vargas, Siri Heidweiller, Márton Géczi, Marcel Kempers, Roland Schmehl*

Faculty of Aerospace Engineering, Delft University of Technology

*Corresponding author. E-mail address: r.schmehl@tudelft.nl

**Abstract**

Generating renewable energy on Mars is technologically challenging. Firstly, because compared to Earth, key energy resources such as solar and wind are weak as a result of very low atmospheric pressure and low solar irradiation. Secondly, because of the harsh environmental conditions, the required high degree of automation and the exceptional effort and costs to transport material to the planet. Like on Earth, it is crucial to combine complementary resources for an effective renewable energy solution. In this work, we present the result of a design synthesis exercise, a 10 kW microgrid solution, based on a pumping kite power system and photovoltaic solar modules to power the construction as well as the subsequent use of a Mars habitat. To buffer unavoidable energy fluctuations and balance seasonal and diurnal resource variations, the two energy systems are combined with a compressed gas storage system and lithium-sulfur batteries. The airborne wind energy solution was selected because of its low weight-to-wing-surface-area ratio, compact packing volume and high capacity factor which enables it to endure strong dust storms in an airborne parking mode. The surface area of the membrane wing is 50 m$^2$ and the mass of the entire system, including the kite control unit and ground station, is 290 kg. The performance of the microgrid is assessed by computational simulation using available resource data for a chosen deployment location on Mars. The projected costs of the system are €8.95 million, excluding transportation to Mars.

## 1. Introduction

Several governmental space agencies and private corporations have proposed human missions to Mars with the goal to establish a habitat. Because such missions are extremely costly and transportation capacity is limited, in situ material and energy utilization will be crucial (Horneck et al, 2003). The environmental conditions on Mars and by that also the renewable energy sources differ substantially from those on Earth. For example, surface temperatures range from -140 to 30 degrees Celsius, surface pressure is more than 100 times less compared to Earth, and the solar irradiance is only 43 percent of that on Earth (Williams, 2020). Solar irradiance is reduced further by strong seasonal dust storms (Fraser, 2009). Wind speeds, on the other hand, can be higher than on Earth. At the Viking land sites, the wind speeds varied between 2 to 7 m/s during summer, 5 to 10 m/s during fall and 17 to 30 m/s during dust storms, resulting in an average of 10 m/s (Boumis, 2017). The main impeding effect of the dust storms is the effect on the irradiation on solar panels, rather than the corrosive effect (Mersmann, 2015). As on Earth, natural energy sources are varying on shorter (minute, hour and day) and longer (year) time scales. For that reason, any renewable energy system on Mars would ideally use complementary types of energy and combine these with energy storage solutions. Yet, it is not only these resource characteristics that make energy harvesting on Mars challenging, but also the required high degree of automation and the aggressive environment, created by the dust storms, the intense cosmic ionising and solar radiation, the extreme temperatures and associated fluctuations.

The use of wind energy on Mars was analyzed by Haslach (1989), who concluded that despite the low density of the Martian atmosphere, wind speeds are high enough to make wind energy competitive with nuclear power in terms of power produced per unit mass. He proposed the design of a mobile and lightweight vertical axis wind turbine, but also identified the variability of the wind as a critical point that would deserve further attention. An extensive study on utilizing local material and energy sources for a Martian outpost was presented by James et al (1998), covering geothermal, solar and wind resources and confirming that the exploitation of wind power would be feasible. Bluck (2001) investigated the combined use of solar and wind energy systems to power a sustainable Mars base, suggesting to use modified cold-weather wind turbines to cover for the missing solar power during month-long Martian global dust storms. Delgado-Bonal et al (2016) on the other hand, came to the conclusion that the use of wind power on Mars is ineffective, due to too low wind speeds. A lightweight horizontal axis wind turbine was proposed by Holstein-Rathlou et al (2018) and tested in a wind tunnel in a simulated Martian atmospheric environment. A very similar, rapidly-deployable turbine concept is developed commercially by the Danish start-up KiteX, promising a power-to-mass ratio of 75 W/kg at a wind speed of 7.5 m/s (KiteX, 2020).

A further reduction of the structural mass required for wind energy harvesting can be achieved by airborne wind energy systems. The innovative technology is based on tethered flying devices, either combining onboard wind turbines with a conducting tether, or converting the pulling power of the flying devices with ground-based generators (Schmehl, 2019). Another advantage next to the reduced mass is that the flight operation of the systems can be adjusted continuously to the available wind resource, by which the capacity factor can be maximized for a given wind profile (Bechtle et al, 2019). Major advances in automatic flight control over the past two decades have contributed to the commercial development of airborne wind energy for terrestrial applications (Vermillion et al, 2020) and a number of different implementations have reached the prototype stage (Nelson, 2020). Within this context, researchers of NASA have proposed to use kites for wind energy harvesting on Mars (Silberg, 2012). A factor supporting the flight operation of airborne wind energy systems on Mars, where aerodynamic forces are generally much lower than on Earth, is the lower gravity.

In this paper, we present a hybrid wind-solar energy system to power the construction and subsequent use of a subsurface Mars habitat. Central component of this microgrid is an airborne wind energy system based on a remote-controlled flexible membrane wing that is operated in pumping cycles. The 20 kW demonstrator system that has been developed at TU Delft for terrestrial applications is illustrated in Figure 1.

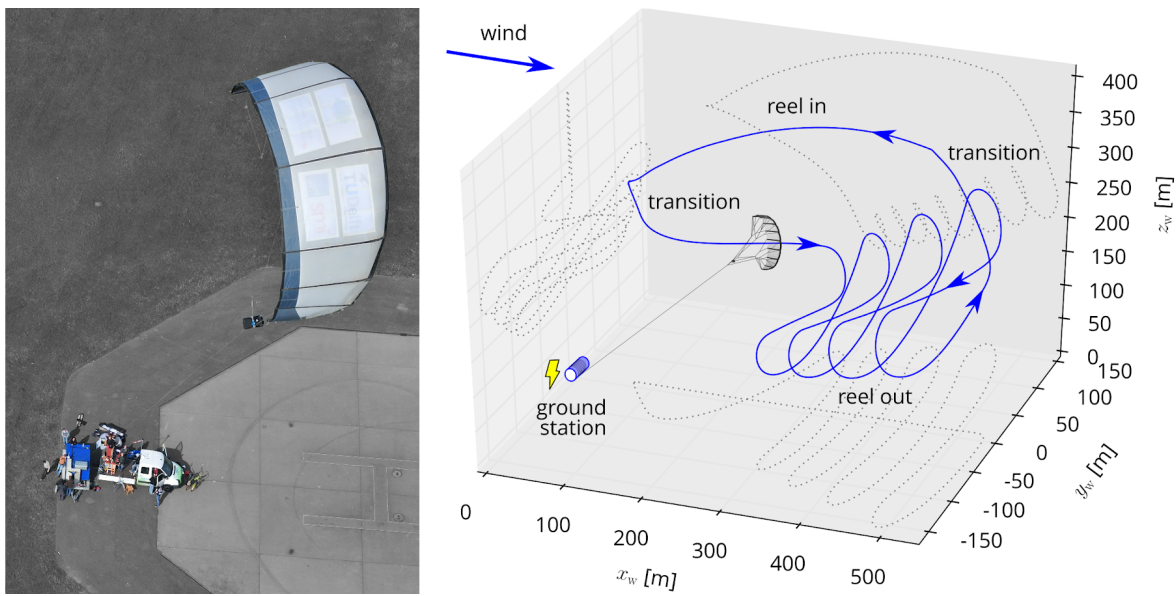

Figure 1: Kite power system of TU Delft in operation and simulated pumping cycle (Fechner, 2016).

During transport, the inflatable lightweight wing can be packed into a compact volume. The generator has a nominal power of 18 kW and produces an average electrical power of about 7 kW in good wind conditions (Van der Vlugt, 2013). This power is sufficient for about 14 Dutch households. To adapt the system to the lower atmospheric density on Mars, the wing surface area is increased. The sizing and design of the kite power system is based on a validated performance model (Van der Vlugt, 2019). The complete microgrid solution also includes photovoltaic (PV) modules and electrical storage to buffer periods of low wind.

The paper summarizes the results of a design synthesis exercise at the Faculty of Aerospace Engineering of TU Delft (Corte Vargas et al, 2020), which succeeded the submission of a proposal to an ESA ideas competition (Bier, 2019). The key contribution of this work is a performance analysis of the combined airborne wind energy and photovoltaic renewable energy solution, accounting for the demand and resource profiles and the coupling of all subsystems. Because only little quantitative information about the energy consumption of the robotic construction of the habitat was available at the time of the study, the design of the microgrid is covering mainly the use of the habitat. It is planned to also integrate the construction phase in a future study. The paper is structured as follows. First, the site for the habitat is selected and the entire mission outlined. Then, the energy system architecture and energy demand of the habitat are defined. Following, the performance of the energy system is analyzed and the design of the microgrid concretized. Subsequent sections detail the sizing of the airborne wind energy subsystem, the solar energy subsystem and the storage solution. Finally, the power management and distribution system are described, followed by conclusions.

**2. Site and Mission**

The siting of the habitat needs to account for several requirements. For one, the basic requirements for a manned mission have to be fulfilled, such as the availability of water and a maximum allowed elevation for landing of -1000 m MOLA (Mars Orbiter Laser Altimeter) (Chen, 2014). This elevation limit ensures a sufficiently high aerodynamic drag used for the braking of the entry module. In addition, the underground has to be suitable for the excavation of the subsurface structures of the planned habitat. Finally, the site must also provide for reasonably good renewable energy sources. Areas on Mars where useful resources for an Earth-independent human settlement on Mars are present, had been explored systematically already in previous research, for example, by James et al (1998). For this project, the site selection started from seven sites chosen from the first round of selection of the NASA landing site selection workshops (NASA, 2020) and from the known SpaceX landing site candidates for its upcoming Mars Starship mission (Foust, 2017). Key selection criteria were the availability of water ice, low elevation, and the possibility of in situ resource utilization (ISRU), with the presence of usable water being the most crucial one. Based on this preselection, a detailed trade-off was conducted and the Dichotomy Boundary Deuteronilus Mensae (DBDM) exploration zone (EZ) identified as the most suitable landing site. This circular area with a radius of 100 km and centered at 39.11°N 23.199°E is located at the south border of the Deuteronilus Mensae region (Head et al, 2015). It is at an elevation of -4000 m MOLA, surrounded by cliffs and ridges at higher elevations, and possibly the remains of a northern low-lands ocean. The terrain features expose all three major geological eras including various rock samples, climate and possible glacial ice history. Continuous water ice is located in high concentrations 10-15 m below the surface (Plaut, 2009) and there is an abundant supply of materials that can be used for ISRU and in-space manufacturing (ISM) for building the habitat (Rummel et al, 2014). Even though solar and wind resources are more scarce in the northern polar region than around the equator, it was concluded that the presence of water and a low elevation were key criteria for the habitat.

The mission starts by launching and deploying the cargo needed for the construction of the habitat. In addition, the servicing orbiter and the Mars Ascent Vehicle (MAV) are sent to Mars. The main goal of the orbiter is to produce, assemble, and perform maintenance on larger-than-payload structures, as well as serving as a gateway between Earth and Mars. The MAV is designed for traveling repeatedly from the Martian surface to orbit and back. Following this preparation phase, the first crew is landing to initialize the deployment and installation of the energy system. Once this system is generating sufficient energy, the construction and subsequent operation of the habitat is started. The energy system will then provide renewable energy to the habitat for a lifespan of 5 Martian years throughout the different seasons, varying from the use of wind and solar energy as the main source of energy. Finally, after the end-of-life of the energy system has been reached, the system can be retired and its different components can be disposed of either on Mars or by sending them back to Earth.

**3. Energy System Architecture and Performance**

The hybrid architecture of the energy system is illustrated in Figure 2, highlighting the connections and interfaces between the subsystems and the external links to the environment, habitat and maintenance system. The five main components are the power management system, the energy storage system, the central control system and the two energy generation subsystems. The primary generation subsystem is based on wind energy, using a flying kite to convert the kinetic energy of the wind into a resultant aerodynamic force and corresponding tether force, which is further converted by the ground-based reeling mechanism and connected generator into shaft power and electrical power, respectively (see Figure 1). The subsystem is equipped with its own control unit and super-capacitor, to balance the energy production and consumption phases of the pumping cycles. Solar PV technology is considered as a secondary generation subsystem, having been used successfully for earlier Mars missions (Delgado-Bonal et al, 2016). The subsystem is equipped with dust protection and tilting mechanisms to minimize losses and ensure the best incidence angle of the radiation. The energy storage system includes short-term storage, using lithium-sulfur batteries to cover the nights, and long-term storage, using $CO_2$ compressed into underground cavities, to cover months with lower resource availability. Carbon dioxide makes up roughly 95% of the Martian atmosphere and the analysis showed that its use for compressed air energy storage (CAES) was meeting the long-term energy storage requirements of the mission. The power management system handles the energy flows from the generation subsystems to the storage system and the habitat. The various configurations of the power distribution subsystem are evaluated; the existing options are traded-off which results in a direct current (DC) microgrid as the most suitable option. The electrical infrastructure is elaborated in section 3.5. The central control system manages the communication of all system components, ensuring the proper functioning of all components.

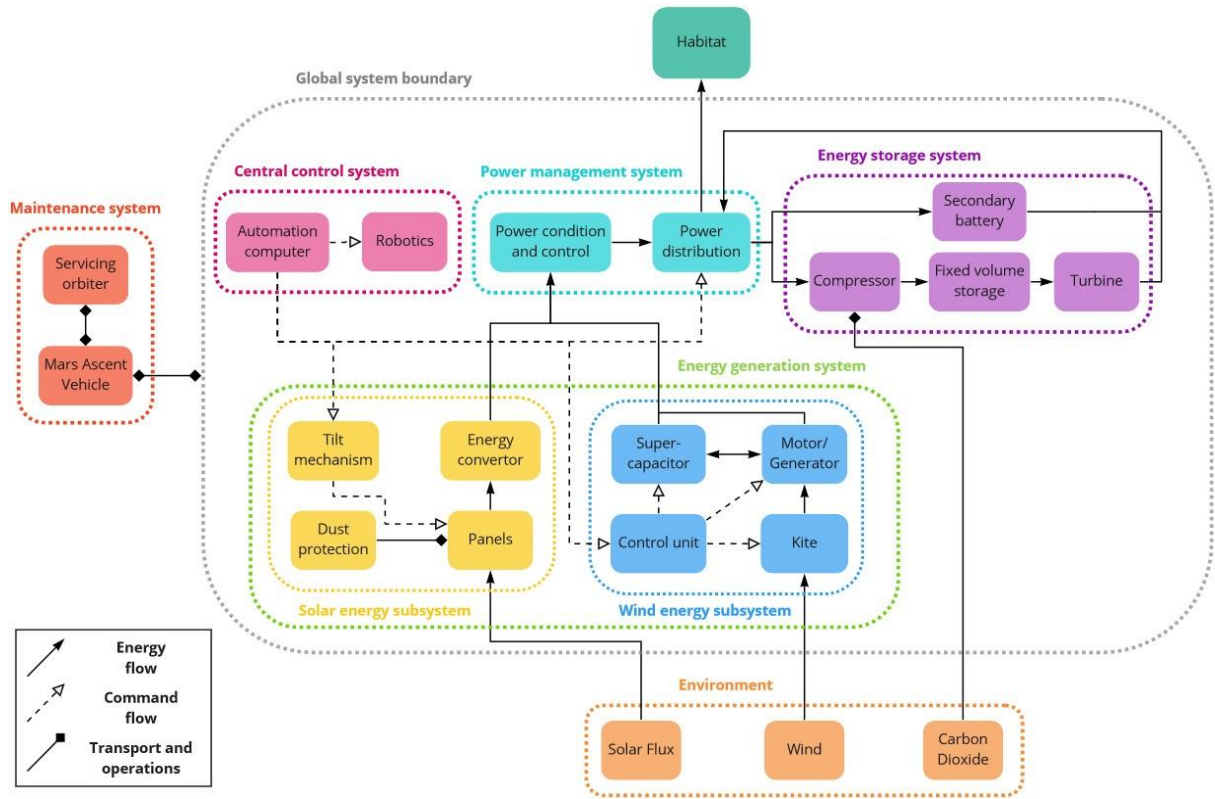

Figure 2: Architecture and interfacing of the entire renewable energy system. Schematic from Corte Vargas et al (2020).

The design of the energy system depends on the energy demand profile of the habitat, the performance of the microgrid including the conversion performances of the PV modules and the kite power system, as well as the availability of the respective energy resources. To determine the energy demand of the habitat, we note that the electrical energy consumption of humans is generally irregular as a result of varying daily activities. Previous research indicates that the mental and physical health of the crew is best maintained throughout the entire mission by adhering to a regular schedule, which helps the human circadian rhythms to stay synchronized with the day-night cycle on Earth (Horneck et al, 2003; Basner, 2013) To aid human adaptation and to ease time keeping, a time-keeping scheme with days of 24 equally lasting hours is considered. Every hour consists of 60 equally lasting minutes and each minute is also composed of 60 seconds. However, a second on Mars is approximately 1.0275 times longer than on Earth. Work and rest periods with different power consumption are distinguished. While work periods require an average of 10 kW for scientific and domestic activities on top of the basic energy requirement of the habitat, rest periods require a lower power for maintaining basic habitat activities and covering the base loads of the life-support systems. Examining and interpreting data from experiments in the Hawaii Space Exploration Analog and Simulation (HI-SEAS) habitat (Barnard et al, 2019; Engler, 2017; Engler et al, 2019), we propose a schedule alternating between 14 martian work hours, lasting from 9am until 11pm, and 10 rest hours with energy requirements of 10 kW and 5 kW, respectively. This leads to a constant daily energy demand of 190 kWh[1].

## 3.1 Performance Model

With the known energy demand of the habitat, the design of the energy system can be determined in multiple iterations between the actual design specification of the five subsystems and the expected performance levels to meet this demand. The process also yields how much of the generated energy is directly used to cover the demand and how much of it is used for short- and long-term storage to cover later parts of the demand. This performance analysis is based on a computational model of the energy system, which is programmed in Python, and which accounts for the performance characteristics and efficiencies of all system components included in Figure 2. The iterative procedure is required to account for the various interdependencies between system components. For example, the microgrid efficiencies depend on the nominal power fed into the grid, while the total energy and power generation depend on the microgrid efficiencies. Also, the outputs of the system model are inputs for the sizing of the subsystems, which is described in the following sections, and vice versa, the results of the sizing are inputs for the performance analysis.

The computational model is implemented in six successive steps (Corte Vargas et al, 2020):

1. Define daily operational conditions
2. Calculate direct supply conditions
3. Evaluate resource conditions
4. Evaluate wind, solar, battery and CAES performance
5. Evaluate power generation for seasonal storage
6. Evaluate power mix values, total energy and nominal powers

The results of step 1 and 2 are illustrated in Figure 3 (left) which shows the astronauts schedule and the resource availability over a complete Martian year. Solar energy is available between sunrise and sunset (Delgado-Bonal et al, 2016), while the availability of the wind resource is determined by the diurnal cycle, with average wind speeds generally dropping during night (Read et al, 2015; Viúdez-Moreiras et al, 2020). Because atmospheric circulation is caused by the differential heating between the equator and the poles and the rotation of the planet, we assume, as a first approximation, that wind energy can be harvested from one hour before sunrise until one hour after sunset for spring, summer and autumn, while during winter it can be harvested from two hours before sunrise until two hours after sunset. The severely limited direct energy supply during nighttime requires that a considerable fraction of the energy during the nights has to be supplied by short-term storage. Thus, during wind and solar harvesting periods, sufficient energy both for the daytime demand and for battery charging needs to be generated.

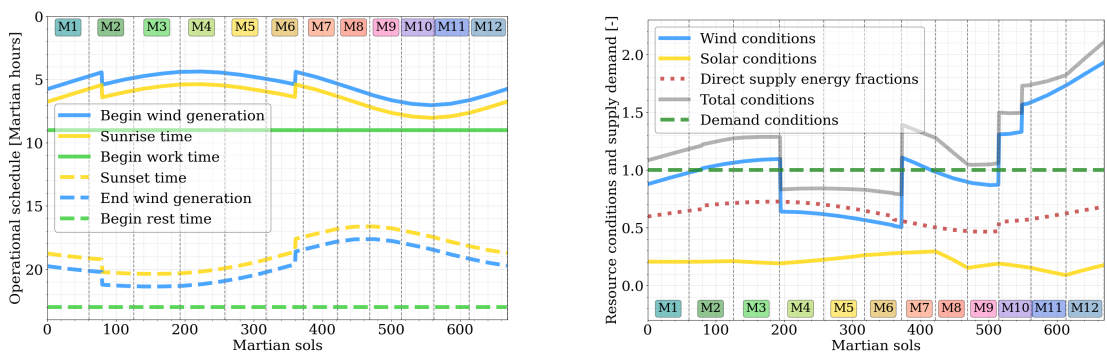

Figure 3: Yearly operational schedule (left) and energy conditions (right). A sol is a solar day on Mars, spanning 24 Martian hours, while one year on Mars lasts 669 sols. In the graphs the twelve Martian months corresponding

---

[1] Energy units kWh and mWh refer to Martian hours

to a solar longitude change of 30 degrees each, are distinguished through the vertical dashed lines and their number given in the labels as adapted from the Mars Climate Database (2006). Every three months the season changes, starting with the northern hemisphere spring equinox at a solar longitude of 0 degrees and Sol 1. Diagrams adapted from Corte Vargas et al (2020).

In step 3, we evaluate the normalized wind and solar energy available, which we denote as wind and solar conditions, or, more generally, as resource conditions. The wind conditions are determined by dividing the daily energy production of the kite by the daily energy demand of the habitat. The demand ranges from 231.5 to 246.4 kWh, accounting for the base demand (190 kWh) and losses due to seasonal fluctuations as well as iterated supply path and subsystem losses. A value of 1 means that the wind energy fully meets the daily demand. The evolution of the wind conditions is obtained from the resource conditions and supply path specifications while its magnitude is ruled by the size of the kite.

Similarly, for the evolution of the solar conditions, the daily irradiance is normalized by the maximum irradiance in the year and divided by a factor representing the averaged supply path losses of that sol. As solar is the secondary/complementary source, the magnitude of the conditions is iterated upon during the design through manually changing the maximum solar condition (resulting to a final value of 0.29). The computed resource conditions are illustrated in Figure 3 (right). Based on the energy conditions, the distribution process of the power management system is simulated considering the dominant and auxiliary source and how they compare to the direct energy supply fraction. For the resulting system design, wind energy is considered to be the dominant source throughout the entire year, because for our design wind conditions are always superior to the solar conditions. The dominant source is used for direct supply as this minimizes losses. If the dominant source is greater than one, the excess energy is used to charge the battery (months 2, 3, 7, and 10 to12). If it is greater than the direct energy fraction but lesser than one, the auxiliary source (solar energy) is also used to charge the battery for night-time energy consumption (months 1, 8 and 9). Any excess energy generated in those months is redirected to the CAES facility for storage purposes. Furthermore, in summer when the sum of the conditions is less than one (months 4 to 6), the CAES facility is used to compensate for the lack of generation.

Lastly, the design is iterated by changing the magnitude of the wind condition, which depends on the kite size, and complementary solar conditions, which depend on the prescribed maximum condition at the end of month 7 as an input. During the iterations, steps 3-6 are repeated with the new input. The nominal power flow in the microgrid is monitored and path losses altered, if necessary. The charging cycles of the CAES facility are examined to guarantee yearly net positive charge, to evaluate the required capacity, and estimate the feasibility of the design. On the same note, the energy required based on the system analysis is compared to the energy generation by the wind and solar subsystems as illustrated in Figure 4, which shows the differences between actually generated and demanded values as modeling errors. A breakdown of the yearly energy flow is illustrated by the Sankey diagram in Figure 5 .

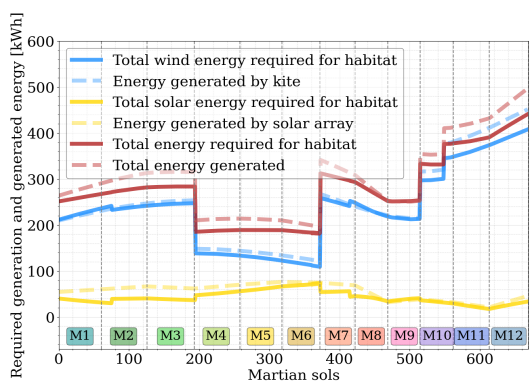

Figure 4: Wind energy required by the system performance analysis and actual output of the designed kite, solar energy required by the system performance analysis and actual energy output of the designed PV array, and total energy required and generated. Diagram adapted from Corte Vargas et al (2020).

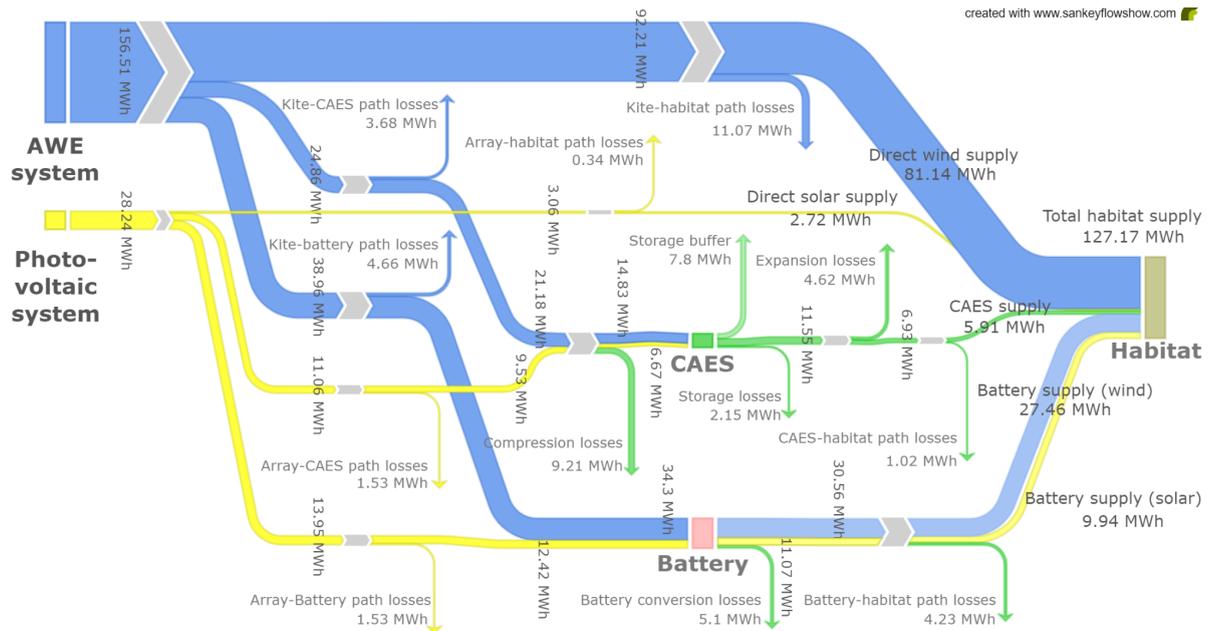

Figure 5: Sankey diagram of total yearly energy generation, losses and supply in Martian MWh.

The resulting annual energy mix fractions computed from the annual supply demand (669 sols x 190 kWh = 127.17 MWh) are summarized in Table 1. In conclusion, the required rechargeable battery capacity is 116 kWh, the microgrid should be able to facilitate a power input of 26 kW, all with a contingency of 5%. The required supply through the CAES is 6.5 MWh with a contingency of 10%. Furthermore, from the net charge of the CAES it could be derived that the mission should be started in autumn/winter to save up enough energy for the summer supply.

| Direct wind | Direct solar | Through battery | Through CAES | Trough wind | Through solar | Through CAES |
|---|---|---|---|---|---|---|
| 81.14 MWh | 2.72 MWh | 37.4 MWh | 5.91 MWh | 108.6 MWh | 12.66 MWh | 5.91 MWh |
| 63.84 % | 2.14 % | 29.37 % | 4.65% | 85.44 % | 9.91 % | 4.65 % |

Table 1: Total annual energy and energy mix for direct and through-storage supply (left) and for daily and seasonal supply (right).

### 3.2 Airborne Wind Energy Subsystem

The choice of a pumping kite power system with a flexible membrane wing was the result of an initial trade-off analysis in which the suitability of different wind energy technologies was assessed for the planned space mission. Horizontal-axis and vertical-axis wind turbine variants were ruled out because of their prohibitive mass and volume. The low specific mass and transportation volume of tethered flying systems makes them generally an interesting option for wind energy harvesting on other planets with an atmosphere. Especially inflatable wings with rigid reinforcements can be scaled up efficiently to compensate for the substantially lower atmospheric density on Mars (Breuer & Luchsinger, 2010). An example for the effect of the lower atmospheric density and gravity on Mars is the aerodynamic design of NASA's Ingenuity drone (NASA, 2020; Chi, 2020).

As illustrated in Figure 1, the pumping kite power system is operated in alternating energy-generating tether reel-out and and energy-consuming reel-in phases. When reeling out, the kite performs cross-wind maneuvers to maximize the flight speed and pulling force. When reeling in, the crosswind maneuvers are discontinued and the kite is depowered, to minimize the pulling force and retraction time. The inflatable wing is actuated by a kite control unit (KCU). This suspended cable robot is responsible for steering the wing and changing its angle of attack, to modulate the pulling force (Van

der Vlugt, 2013). A flight path planner adjusts the planned path to the instantaneous wind conditions and a flight controller uses sensor data from kite and ground station to ensure that the kite follows the planned path (Jehle & Schmehl, 2012; Fechner & Schmehl, 2018; Rapp & Schmehl, 2019). The ground station is responsible for controlling the tether force and thereby optimize the generation of power (Fechner & Schmehl, 2013). The functional architecture of the pumping kite power system adapted to operation on Mars is shown in Figure 6.

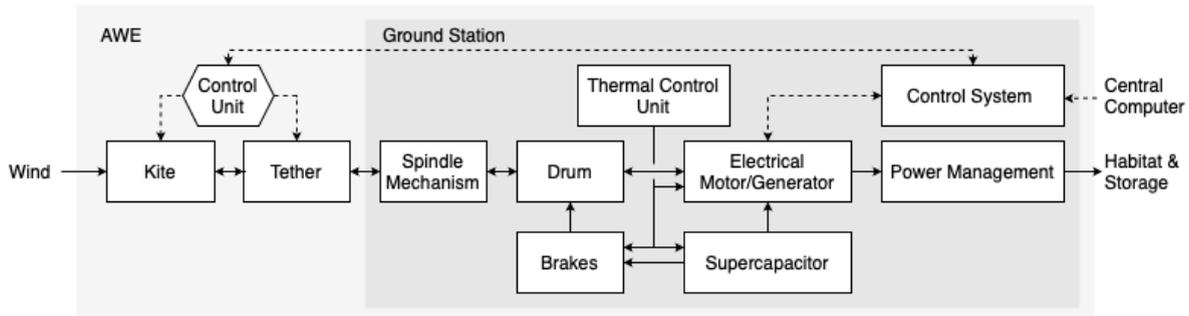

Figure 6: AWE system architecture for operation on Mars. Solid lines denote energy flows, dotted lines communication flows. Schematic from Corte Vargas et al (2020).

To size the AWE system as the primary energy supplier of the microgrid, a computational model was developed in Python, accounting for the main aspects of the system: kite operation (Luchsinger, 2013), axial flux generator (Fechner & Schmehl, 2013) and tether (Bosman et al, 2013). Outputs of the model include the weight of the AWE system along with its energy output throughout the year. The model is embedded in the iterative design procedure, such that the sizing of all subsystems of the microgrid can be done concurrently, while ensuring that all requirements are met. The computational model was verified and validated by comparing the results to a validated performance model of the pumping kite power system (Van der Vlugt et al, 2019; Schelbergen & Schmehl, 2020).

As a result of this analysis, we found that a 50 m$^2$ kite with an average annual power output of 25 kW is needed for sustaining the habitat. The AWE system has a total mass of 290 kg, can be compacted to a volume of 0.7 m$^3$ and costs approximately €70,000. Suitable materials were determined by trade-off analyses for all system components, considering the mass, structural properties and sustainability of the materials. It was found that to retire the AWE system on Mars, many of the components and materials can be reused to minimize the environmental impact.

**3.3 Solar Energy Subsystem**

Because of its high technology readiness level (TRL), reliability and mass performance, solar PV was identified as the most suitable secondary energy technology for the planned mission. To compensate for the low solar irradiance and dust storms on Mars, we selected a dual-axis support system with a sun simulation for sun tracking to reduce the incidence angle on the panels. As for the dust settling on the panels, we propose an experimental coating (Zhou, et al, 2019) that was tested under simulated Mars conditions and is expected to be able to remove more than 90% of all dust from the panels by inclining them, while having a negligible effect on the transparency of the panel cover.

Space-grade solar panels are usually designed for optimal performance at an Air Mass 0 (AM0) spectrum, considering direct light from the sun that has not interacted with an atmosphere (Funde, 2020). Compared to AM0 and the spectrum on the surface of Earth, the spectrum at the surface of Mars differs substantially as the shorter wavelengths are filtered out to a large extent by the red dust (Landis & Hyatt, 2006). When using multijunction solar cells – which would be required because of their higher conversion efficiency, leading to lower required volume and mass – the change in spectrum could render the solar cells to no longer be current matched. To counteract this, Landis & Hyatt (2006) found that triple-junction GaInP/GaAs/Ge cells could be adapted to the Mars surface spectrum as its specific band gaps are well suited for the long wavelength-rich spectrum, having an energy conversion efficiency of approximately 32%.

During their operative lifetime, the PV panels will produce no emissions and thus will leave little to no impact on Mars. However, the production of the panels results in waste due to the refinement of rare metals needed for the triple-junction cells. At end of life, most of the subsystems and the panels can be recycled, although specialized facilities are needed to disassemble the panels and purify the metals for reuse. This can not be done on Mars, as it is assumed that the required facilities will not be built in the timespan of the mission. Thus, to reduce the impact on the Mars environment it is advised to return the panels to Earth for recycling at mission end.

A system for the cooling of the PV panels is still to be developed. As the atmosphere is thinner on Mars, the panels do not cool as effectively as they would in an Earth environment and thus the efficiency of the panels is reduced according to an initial estimation by [Delgado-Bonal et al (2016)](). Assuming that an efficient cooling solution can be developed, the area of the solar array required for supplementing the wind energy system would be approximately 70 m$^2$. The complete cost of the solar energy system would amount to an estimated €6,800,000, with a mass of around 790 kg.

### 3.4 Energy storage Subsystems

Apart from production and distribution of the energy, buffering and storage are key aspects of any autarkic renewable energy system. In our case, the produced energy is distributed via three different paths: a part of the energy is directly supplied to the habitat, while the excess part is captured by rechargeable batteries, for short-term storage covering the nights, and compressed air energy storage (CAES), for covering seasonal generation shortages. An initial study led to a required cavern volume and pressure ratio of 107500 m$^3$ and 175, respectively. The mass of $CO_2$ to be stored annually is 301812 kg, roughly 86% of the total storage system weight, disregarding the regolith that is mined and used as inner structural lining material. Because of mechanical, thermal and electrical losses of the cycle components (expander, compressor, convertors), the overall efficiency of the CAES system is around 45%. The cavern volume is limited by the maximum stresses that the rock-mass formation of Martian terrain can handle. Further investigations into the local rock quality is recommended to provide more insight to the maximum storage capacity and required distance from the habitat, to reduce the operational risk. The costs for the CAES system is estimated to be €2,000,000 for development, transport and subsystem-specific mission operations.

Although lithium-ion batteries have become a standard in the space industry, lithium-sulfur batteries were chosen for the systems' day-to-day storage, because these batteries are well recyclable ([Deng, 2017]()) and their energy density (350 Wh/l) is higher than that of lithium-ion batteries (200 Wh/l). The batteries are charged during the day and discharged during the night. A volume of 0.1 m$^3$ and a mass of 120.4 kg were determined during the system performance analysis, based on the required capacity needed to get through the longest nights on Mars ([Wertz, 1999]()). The batteries will have to be stored in a protective container shielded from Mars' environment and they will be modular to allow for easy replacement or repairs. Finally, the cost of the batteries will end up at an estimated €6450.

### 3.5 Power Management and Distribution System

To ensure a reliable delivery of the electrical power to the habitat, where it is consumed, a well-engineered power management and distribution system is critical. This is achieved by a direct current microgrid, which interconnects the habitat and the various energy generation and storage systems. A DC microgrid is preferred to the more widely applied alternating current (AC) alternative, because this grid type requires less power conversion devices, which is favorable in terms of overall mass, volume, efficiency and reliability. The critical system elements of the microgrid are schematically illustrated in [Figure 7]().

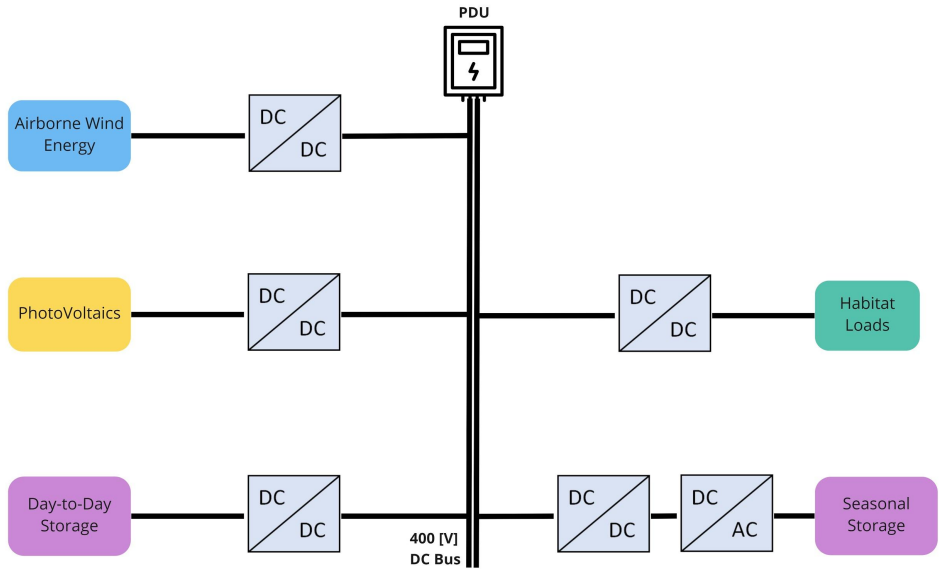

Figure 7: Schematic of the microgrid system architecture, adapted from Corte Vargas et al (2020).

Power electronic converters are used to convert the electrical power that is injected into (or withdrawn from) the microgrid to the correct current and voltage levels. State-of-the-art Silicon-carbide (SiC) converters for aerospace applications were selected, to facilitate high power densities and minimize power losses. A power distribution unit (PDU) is used for monitoring and controlling the power flow in the microgrid. Lastly, the DC bus is the common connection to which all power sources and sinks are connected in parallel. The power cables are laid under the Martian surface for better protection against the harsh environment. The grid voltage is maintained at 400 VDC, which has been determined to be the optimum voltage according to a study on voltage levels for DC microgrids (Anand & Fernandes, 2010).

Altogether, the power management and distribution system is designed to support a nominal power of 26 kW. The electrical path efficiencies, which indicate the power losses from supply to demand, are computed to be in the range of 73 to 89%. The total mass of the system is estimated at 200 kg, for a volume of 0.24 m$^3$ and an approximate cost of €70,000.

### 4. Conclusion

The harsh environmental conditions on Mars due to dust storm seasons, low solar irradiation and low atmospheric pressure, require specific technical solutions. As for any space mission, a high degree of automation is required and the effort and costs to transport material have to be minimized. These requirements converge in the goal to design an energy system for a Mars habitat that is sustainable, lightweight, as low in costs as possible while reliably fulfilling the energy requirements.

To overcome the seasonal, monthly and daily fluctuations in sustainable energy generation on Mars, the energy system is based on two different renewable energy technologies, a storage system and a power management system. The primary energy system consists of a pumping kite power system with a semi-rigid inflatable kite. Due to the lightweight materials and large wing surface area, the kite system is able to perform at the low Martian atmospheric density. The lightweight materials and high foldability make it suitable for transportation to Mars. The secondary energy system uses solar PV panels with a dual axis-system support system. The panels follow the tracking of the sun, to reduce the incidence angle to maximize the yield. To remove dust pollution of the PV panels, a coating is applied which removes over 90% of the dust by inclination of the angles, without reducing its transparency. In addition, the selected cells are able to perform at the long wavelength-rich Martian spectrum.

As both wind and solar energy are fluctuating resources, an energy storage system is designed to which excess energy is charged during harvesting times. It consists of a day-to-day lithium-sulfur battery, which provides energy to the habitat overnight, and a seasonal compressed air storage system, to overcome seasons where less energy can be produced. The latter makes use of a local cavity, reducing the transportation volume. Finally, the power management system connects the energy harvesting systems and the storage solutions, ensuring reliable electric delivery to the habitat. It makes use of a DC microgrid with underground power cables, to protect against the harsh Martian conditions.

The combination of all these subsystems results in a design that can reliably produce and distribute enough energy for the Mars habitat, at a total base cost of €8.95 million, excluding transportation. This proves that renewable energy is a feasible option for a Mars mission and that further investigation needs to be done to finalize the design. For future work, we recommend to determine the environmental conditions on Mars more accurately, in particular the wind resource, and to refine the system model and further detail the individual components of the microgrid. Moreover, from the price difference of the space-grade photovoltaic subsystem and the estimated cost of a space-grade AWE subsystem, it is evident that further research must be conducted to obtain a more practical expenditure prognosis for the kite and remaining subsystems.

**Acknowledgements**

The authors would like to acknowledge the support of the coaches Dominic von Terzi, Botchu Jyoti and Camila Brito, as well as the experts Angelo Cervone, Henriette Bier, Mark Schelbergen, Mihir Metha, Klaas Akkerman, Joep Breuer and Henk Polinder during the design synthesis exercise.